# Qualitative Comparison of Community Detection Algorithms

Günce Keziban Orman[1,2], Vincent Labatut[1] and Hocine Cherifi[2]

[1] Galatasaray University, Computer Science Department, Istanbul, Turkey
[2] University of Burgundy, LE2I UMR CNRS 5158, Dijon, France
korman@gsu.edu.tr

**Abstract.** Community detection is a very active field in complex networks analysis, consisting in identifying groups of nodes more densely interconnected relatively to the rest of the network. The existing algorithms are usually tested and compared on real-world and artificial networks, their performance being assessed through some partition similarity measure. However, artificial networks realism can be questioned, and the appropriateness of those measures is not obvious. In this study, we take advantage of recent advances concerning the characterization of community structures to tackle these questions. We first generate networks thanks to the most realistic model available to date. Their analysis reveals they display only some of the properties observed in real-world community structures. We then apply five community detection algorithms on these networks and find out the performance assessed quantitatively does not necessarily agree with a qualitative analysis of the identified communities. It therefore seems both approaches should be applied to perform a relevant comparison of the algorithms.

## 1 Introduction

The use of networks as modeling tools has spread through many application fields during the last decades: biology, sociology, physics, computer science, communication, etc. (see [1] for a very complete review of applied studies). Once a system has been modeled, the resulting network can be analyzed or visualized thanks to some of the many tools designed for graph mining. Such large real-world networks are characterized by a heterogeneous structure, leading to specific properties. In particular, a heterogeneous distribution of links often results in the presence of a so-called community structure [2]. A community roughly corresponds to a group of nodes more densely interconnected, relatively to the rest of the network [3]. The way such a structure can be interpreted is obviously dependent on the modeled system. However, independently from the nature of this system, it is clear the community structure conveys some very important information, necessary to a proper understanding [4]. Detecting communities is therefore an essential part of modern network analysis.

Many different community detection algorithms exist [2], which can diverge in two ways: first the process leading to an estimation of the community structure, but also the nature of the estimated communities themselves. This raises a question regarding the comparison of these algorithms, from both a theoretical and a practical point of view. Authors traditionally test their community detection algorithms on real-world and/or artificial networks [5, 6]. The performance is assessed by comparing the estimated communities with some community structure of reference using a quality (e.g. modularity [3]) or association (e.g. Normalized Mutual Information [7]) measure. This single value is then compared with those obtained when applying preexisting algorithms on the same data. The main problem with this approach is its purely quantitative aspect: it completely ignores the nature of the considered community structures. Two algorithms can reach the exact same level of performance, but still estimate very different community structures.

Besides performance evaluation, the data used to perform the tests are also subject to some limitations. For realism reasons, it is necessary to apply the algorithms to real-world networks, but these are not sufficient because 1) reference communities (i.e. ground truth) can rarely be defined objectively, and 2) the topology of the selected networks can hardly be diverse enough to represent all types of systems. Testing on artificial networks can be seen as complementary, because they overcome these limitations. Indeed, a random model allows generating as many networks as desired, while controlling some of their topological properties. The only issue is the realism of the obtained networks, which should mimic closely real-world networks in order to get relevant test results.

Some properties common to most real-world networks are well-identified: power-law distributed degree, small-worldness, non-zero degree correlation and relatively high transitivity [8]. Additionally, networks with a community structure are characterized by a power-law distributed community size [9]. Several generative models with increasing realism were successively designed [6] before finally meeting these constraints [6, 10, 11]. However, recent studies showed other properties besides the community size distribution can be used to characterize real-world community structures [4, 12], such as community-wise density and average distance, hub dominance, embeddedness. They allow giving a more detailed description of the internal topology of the communities, and of the way they are interconnected.

To our opinion, these new results have two important consequences on the problem of community detection assessment. First, the question to know whether the artificial networks used as benchmark also exhibit these properties arises naturally. But more importantly, it is now possible to perform a qualitative comparison of the communities identified by different algorithms, instead of relying only on a single performance measure. In this article, we try to answer both questions using the realistic generative model LFR [6] and a representative set of community detection algorithms. In section 2, we review in greater details the properties used to describe community structures in complex networks. In section 3, we shortly present the LFR model and its properties, and introduce an adjustment allowing to improve the realism of the networks it generates. We also review the various approaches used in the literature to define the concept of community, and describe a selection of community detection algorithms from this perspective. In section 4, we first analyze the properties of the generated networks. We then compare the community structures detected by the selected algorithms on these networks. We consider both their similarity to the community structure of interest, and how community properties differ from one algorithm to the other. Finally, we discuss our results and explain how our work could be extended.

## 2      Characterization of a Community Structure

Complex networks are often characterized at microscopic and macroscopic levels, i.e. by studying the characteristics of nodes taken individually and of the network considered as a whole, respectively. The microscopic approach focuses on some nodes of interest, and tries to identify which features allow distinguishing them from the rest of the network (degree, centrality, local transitivity, etc.). At the macroscopic level, one can take advantage of the multiplicity of nodes to derive statistics or distributions summarizing some of the network features (degree distribution, degree correlation, average distance, transitivity, etc.). The development of community detection corresponds to the apparition of a mesoscopic level, and highlights the need for adapted tools to characterize the community structure. In this section, we present a selection of the mesoscopic measures recently proposed and indicate how real-world networks behave relatively to them. In some cases, no general observation can be made, and one has to consider the class of the network: communication, biological, social, etc. Note other measures exist besides those described here, such as the network community profile [12] or roles distribution [13].

The *community size distribution* is considered as an important characteristic of the community structure. It has been largely studied in real-world networks, and seems to follow a power law [9, 14] with exponent $\beta$ ranging from 1 to 2 [15]. This means their sizes are heterogeneous, with many small communities and only a few very large ones.

The *embeddedness* measure assesses how much the direct neighbors of a node belong to its own community. It is defined as the ratio of the internal degree $k_{int}$ to the total degree $k$ of the considered node [4]:

$$e = k_{int}/k \qquad (1)$$

This internal degree is the number of links the nodes has with other nodes from the same community, by opposition to its external degree $k_{ext}$, which corresponds to connections with nodes located in other communities. The maximal embeddedness of 1 is reached when all the neighbors are in its community ($k_{int} = k$), whereas the minimal value of 0 corresponds to the case where all neighbors belong to different communities ($k_{int} = 0$). In real-world networks, a majority of nodes, usually with low degree, have a very high embeddedness. The rest are distributed depending on the considered class: communication, Internet and biological networks exhibit a peak around $e = 0.5$, whereas social and information networks have a more uniform distribution. In all cases, the whole range of $e$ is significantly represented, even small values [4].

The *density* $\rho$ of a community $C$ is defined as the ratio of links it actually contains, noted $m_C$, to the number of links it could contain if all its nodes were connected. In the case of an undirected network, the latter is $n_C(n_C - 1)/2$, where $n_C$ is the number of nodes in the community, and we therefore get $\rho = 2m_C/(n_C(n_C - 1))$. When compared to the overall network density, the density allows assessing the cohesion of the community: by definition, a community is supposed to be denser than the network it belongs to. The *scaled density* is a variant obtained by multiplying the density by the community size [4]:

$$\tilde{\rho}(C) = \rho(C)n_C = \frac{2m_C}{n_C - 1} \qquad (2)$$

If the considered community is a tree, it has only $m_C = n_C - 1$ links, and $\tilde{\rho}(C) = 2$. If it is a clique (completely connected subgraph), then $m_C = n_C(n_C - 1)/2$ and we have $\tilde{\rho}(C) = n_C$. The scaled density therefore allows characterizing the structure of the community. Some real-world networks such as the Internet or communication networks have essentially tree-like communities. On the contrary, for other classes like social and information networks, the scaled density increases with the community size. Finally, biological networks exhibit a hybrid behavior, their small communities being tree-like whereas the large ones are denser and close to cliques [4].

The *distance* between two nodes corresponds to the length of their shortest path. When averaged over all pairs of nodes in a community, it allows assessing the cohesion of this community. In real-world networks, small communities ($n_C \leq 10$) are supposedly small-world, which means the average distance $\ell$ should increase logarithmically with the community size $n_C$ [4]. For larger communities, the average distance still increases, but more slowly, or even stabilizes for certain classes like communication networks. A small average distance can be explained by a high density (social), the presence of hubs (communication, Internet), or both (biological, information).

From a community structure perspective, a hub is a node connected to many of the other nodes belonging to the same community. The presence of a central hub in a community $C$ can be assessed using the *hub dominance* measure, which corresponds to the following ratio:

$$h(C) = \max_C(k_{int})/(n_C - 1) \qquad (3)$$

The numerator is the maximal internal degree found in $C$, and the denominator is the maximal degree theoretically possible given the community size. The hub dominance therefore reaches 1 when at least one node is connected to all other nodes in the community. It can be 0 only if no nodes are connected, which is unlikely for a community. In real-world networks, the behavior of this property depends on the considered class. For communication networks, it is close to the maximum for all community sizes, meaning hubs are present in all communities. Considering their communities are sparse and tree-like, one can conclude they are star-shaped. Other classes do not have as many hubs in their large communities, which is why their hub dominance generally decreases with community size increase [4].

## 3    Methods

Our experiment is two-stepped: first we generate a set of artificial networks and study the realism of their community-related topological properties; second we apply a selection of community detection algorithms on these networks and analyze the properties of the community structure they estimate. In this section, we first describe the LFR model we applied during the first step, which supposedly allows generating the most realistic networks in terms of overall properties (degree distribution, small-worldness, etc.) [6, 10, 11]. Then, we shortly describe the community detection algorithms we selected, and explain how they differ on the way they handle the concept of community.

### 3.1    Network Generation

Only a few models have been designed to generate networks possessing a community structure. Girvan and Newman seemingly defined the first one [5], which produces networks taking roughly the form of sets of small interconnected Erdős-Rényi networks [16]. Although widely used to test and compare community detection algorithms, the Girvan-Newman method is limited in terms of realism [6], mainly because the degree is not power-law distributed and the communities are small, few, and even-sized. Several variants were defined, allowing to produce larger networks and communities with heterogeneous sizes [2, 7, 17]. More recently, a different approach appeared, based on a rewiring process [6, 18]. It increased the realism level even more by generating networks with power-law distributed degree. Among these newer models, we selected the LFR model, which seems to be the more realistic and was previously used as a benchmark to compare community detection algorithms [6, 10, 19].

The LFR model was proposed by Lancichinetti *et al*. [6] to randomly generate undirected and unweighted networks with mutually exclusive communities. The model was subsequently extended to generate weighted and/or directed networks, with possibly overlapping communities [19]. However, in this article, we focus on undirected unweighted networks with non-overlapping communities, because the community structure-related properties we want to study have been defined and/or used only for this type of networks. The model allows to control directly the following parameters: number of nodes $n$, desired average $\langle k \rangle$ and maximum $k_{max}$ degrees, exponent $\gamma$ for the degree distribution, exponent $\beta$ for the community size distribution, and mixing coefficient $\mu$. The latter represents the desired average proportion of links between a node and nodes located outside its community, called inter-community links. Consequently, the proportion of intra-community links is $1-\mu$. A node of degree $k$ has therefore an external degree of $k_{ext} = \mu k$ and an internal degree of $k_{int} = (1-\mu)k$.

The generative process first uses the configuration model (CM) [20] to generate a network with average degree $\langle k \rangle$, maximum degree $k_{max}$ and power-law degree distribution with exponent $\gamma$. Second, virtual communities are defined so that their sizes follow a power-law distribution with exponent $\beta$. Each node is randomly affected to a community, provided the community size is greater or equal to the node internal degree. Third, an iterative process takes place to rewire certain links, in order to approximate $\mu$, while preserving the degree distribution. For each node, the total degree is not modified, but the ratio of internal and external links is changed so that the resulting proportion gets close to $\mu$.

By construction, the LFR method guaranties to obtain values considered as realistic [1, 8] for several properties: size of the network, power law distributed degrees and community sizes. Other properties are not directly controlled, but were studied empirically [10]. It turns out LFR generates small-world networks, with relatively high transitivity and degree correlation. This is realistic [8], but holds only under certain conditions. In particular, transitivity and degree correlation are dramatically affected by changes in $\mu$, and become clearly unrealistic when it gets different from 0. An adjustment was proposed to solve this issue, consisting in using a different generative model during the first step [11]. By applying Barabási & Albert's preferential attachment model (BA) [21] instead of the CM, the degree correlation and transitivity become more stable relatively to changes in $\mu$.

It is rather clear the mixing coefficient is complementary to the embeddedness presented in section 2 (eq. 1): $e = 1 - \mu$. Yet, it was mentioned in the same section the embeddedness varies much from one node to the other in real-world networks, exhibiting bimodal and flat distributions. From this point of view, the LFR model is not realistic, since it produces networks whose nodes have all roughly the same mixing coefficient. To solve this problem, we implemented a small adjustment allowing to specify the complete distribution of $\mu$ in place of a single objective value.

### 3.2    Community Detection

Because of their number and great diversity, it is difficult to categorize community detection algorithms. Here, we chose to characterize them not by considering the process they implement, as it is usually done, but rather the definition of the community concept they rely upon. We selected a representative set of algorithms, favoring fast ones because of the size of the analyzed networks. We give very partial descriptions in this section, so the reader might want to consult the review by Fortunato [2] to find more information concerning community detection. A very widespread informal definition of the community concept considers it as a group of nodes densely interconnected compared to the other nodes [2, 14, 22]. In other terms, a community is a cohesive subset clearly separated from the rest of the network. Formal definitions differ in the way they translate and combine both these aspects of cohesion and separation.

A direct translation of the informal definition given above consists in first specifying two distinct measures to assess separately cohesion and separation, and then processing an overall measure by considering their difference or ratio. This approach led to many variants, differing on how the measures are defined and combined. The most widespread one is certainly the modularity, a chance-corrected measure which assesses cohesion and separation through the number of intra- and inter-community links, respectively. We selected two modularity optimization algorithms, which differ in the way they perform this optimization. *Fast Greedy* applies a basic greedy approach [3], and *Louvain* includes a community aggregation step to improve processing on large networks [23].

Another family of approaches is based on node similarity measures. Such a measure allows translating the topological notions of cohesion and separation in terms of intra-community similarity and inter-community dissimilarity. In other terms: a community is viewed as a group of nodes which are similar to each other, but dissimilar from the rest of the network. Once all node-to-node similarities are known, detecting a community structure can be performed by applying a similarity-based classic cluster analysis algorithm [24]. We selected the *Walktrap* algorithm, which uses a similarity based on random walks and applies a hierarchical agglomerative clustering approach [17].

Some approaches based on data compression do not use the cohesion and separation concepts like the previous definitions. They consider the community structure as a set of regularities in the network topology, which can be used to represent the whole network in a more compact way. The best community structure is supposed to be the one maximizing compactness while minimizing information loss. The quality of the representation is assessed through measures derived from mutual information. Algorithms essentially differ in the way they represent the community structure and how they assess the quality of this representation. We selected the *InfoMap* algorithm [25], whose representation is based on coded ids affected to nodes.

The definition of the community concept is not always explicit: procedural approaches exist, in which the notion of community is implicitly defined as the result of the processing. To illustrate this, we selected the *MarkovCluster* algorithm, which simulates a diffusion process in the network to detect communities [26]. This approach relies on the transfer matrix of the network, which describes the transition probabilities for a random walker evolving in this network. Two transformations are iteratively applied on this matrix until convergence. The resulting matrix can be interpreted as the adjacency matrix of a network with disconnected components, which correspond to communities in the original network.

We have to mention another family of algorithms based on link centrality. They iteratively remove the most central links until disconnected components are obtained, which are interpreted as the network communities. The community structure largely depends on the selected centrality measure, e.g. edge-betweenness [5]. However, the computational cost of such algorithms is very high and we were not able to apply them to our data.

## 4      Results and Discussion

### 4.1    Properties of the Generated Communities

Using our review of the literature, we selected realistic values for the parameters the LFR model lets us control. We used three different network sizes: $n \in \{10000, 100000, 500000\}$, constant average and maximal degrees $\langle k \rangle = 30$, $k_{max} = 1000$, and exponent $\gamma = 3$ for the degree power-law distribution. The exponent was $\beta = 2$ for the community size distribution, whose bounds were $k_{min}$ and $k_{max}$. The mixing coefficient $\mu$ was distributed uniformly over its definition domain $[0; 1]$. We generated 5 instances of network for each combination of parameter values, in order to check for consistency.

Among the community-related properties we described in section 2, two are directly controlled by the LFR model: the community size and embeddedness distributions. Our measurements confirm on all networks that the community sizes follow a power-law distribution as expected (cf. Fig. 1). Note the range of these sizes varies much from one real-world network to the other, and it is therefore difficult to describe a typical set of values. However, we can say the communities we generated are very similar in size to those from real-world networks of comparable size. For instance, we have communities containing between 15 and 700 nodes for $n = 10000$, which is compliant with what was observed in networks of this size [4]. For the embeddedness, we obtained a uniform distribution, as expected. It is close enough to what can be observed in social and information networks. The main difference is we do not have as many nodes with very high embeddedness as in those real-world networks. This could be easily corrected though, by specifying a more appropriate distribution when applying the modified LFR model.

We now focus our attention on the uncontrolled properties. The results are very similar independently from the size of the network. The only difference seems to be that values measured on larger networks exhibit slightly smaller dispersion. For this reason we present only results for networks with size $n = 10000$.

The scaled density increases from 11 to 22 along with the community size. This means the smallest communities are clique-like ($\tilde{\rho}(C) = n_C$), and no tree-like communities are generated ($\tilde{\rho}(C) = 2$). These features cannot be considered as realistic: as mentioned before, in real-world networks the small communities are tree-like and the large ones are either tree-like too, or much more clique-like. In other words: real-world networks exhibit two different behaviors, but the generated networks can be compared to none of them. It seems the links are distributed too homogeneously over the generated networks, making small communities too dense and large ones too sparse.

As shown in Fig. 1, the average distance increases regularly from 1.5 to 2.5 along with the community size. The main difference with real-world networks is these have a much lower average distance for smallest communities, reaching values slightly greater than 1. Consequently, we do not observe for the generated networks the fast increase of average distance which was characteristic of the real-world networks. For the rest of the communities, the observed distribution is comparable with communication networks though, with a stable average distance for medium and large communities. Moreover, the values measured for these communities are also realistic in terms of magnitude.

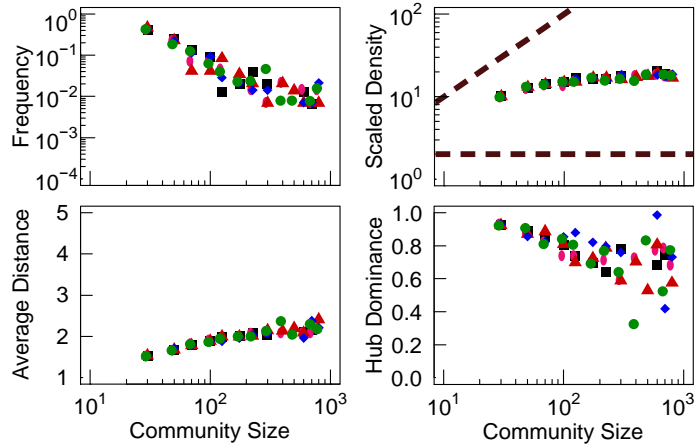

**Fig. 1.** Properties of the generated communities. Each network instance is represented with a different shape/color. Points are averages over logarithmic bins of the community size. The dotted lines in the scaled density plot represents its limits ($\tilde{\rho}(C) = 2$ and $n_C$, cf. section 2).

Hub dominance is very high for the smallest communities, with values close to 1. For large communities, there is no general trend in $n = 10000$ networks: the property varies much over the networks we generated. This dispersion decreases when the network size increases though, and $n = 500000$ networks show a hub dominance decrease with community size increase, reaching values close to 0.3. This behavior is compatible with most classes of real-world networks, which exhibit hubs dominance mainly for small communities. Moreover, the same dispersion was also observed on real-world networks [4]. This measure completely relies on the way high degree nodes are distributed over communities, since it directly depends on the maximal internal degree found in communities. The fact there are much less large communities, due to their power law-distributed sizes, can explain this dispersion. A possible solution would be to consider a measure based on the $k$ highest internal degrees of the community instead of a single one.

To summarize our observations: the generated communities exhibit some, but not all, of the properties observed on real-world networks. Their sizes are realistic, but the distribution of links is not always appropriate. The small communities are too dense and clique-like, when they should be sparser and tree-like, with a smaller average distance. In other terms, they should be star-shaped. They possess the high hub dominance characteristic of such structures though, but this is certainly due to their clique-like configuration. The fact their average distance is much higher than in comparable real-world communities is a surprise. Indeed, one would expect such dense, hub-dominated communities to have a lower average distance. It turns out they are constituted of a clique core and a few very low degree nodes connected to this core: the latter explain the relatively high average distance. The larger communities, on the contrary, should be substantially denser and more clique-like. In some cases, their hub dominance is relatively low despite their small average distance and low density, which seems to indicate they do not contain a main central hub, but several interconnected medium ones. By definition, this feature is not reflected by the hub dominance measure, which only considers the maximal degree in the community.

Another issue is the fact generated networks do not comply with a specific class of real-world networks, but rather have similarities with different classes depending on the considered property. Their average distance have common points with communication networks, whereas this is not the case at all for their embeddedness and hub dominance distributions, which look like social and biological networks. Despites these limitations, the model produces what we think to be the most realistic networks to date, which is why the generated networks constitute an appropriate benchmark to analyze community detection algorithms.

## 4.2    Comparison of the Estimated Communities

We applied the selected community detection algorithms on the generated networks: *Louvain* (LV), *Fast Greedy* (FG), *MarkovCluster* (MC), *InfoMap* (IM) and *Walktrap* (WT). For time matters, it was possible to process networks with sizes 10000 and 100000, but not 500000. We however generated denser $n = 100000$ networks, with $k_{max} = 3000$ (instead of 1000), in order to study the effect of density. Table 1 displays the performance of each algorithm expressed in terms of *Normalized Mutual Information* (NMI), which is a measure assessing the similarity of two partitions (in our case: the reference and estimated community structures). It is considered to be a good performance measure for community detection, and was used in several studies [6, 7, 10]. According to the NMI, IM clearly finds the closest community structure to the reference, followed by MC, LV, and WT, while FG is far behind. This type of quantitative analysis is characteristic of existing works dealing with algorithms comparison. In the rest of this section, we complete it with a qualitative analysis based on the previously presented community properties.

We first focus on the results obtained on $n = 10000$ networks. As we can see on Fig. 2, most algorithms have found communities whose sizes distribution is reminiscent of the power-law used during network generation. However, important differences exist between them. First, MC visibly finds many very small communities ($n_c < 5$), and the other sizes are consequently strongly under-represented. A more thorough verification showed most of these communities are even single nodes, which is particularly problematic since community identification consists in grouping them. It is important to remark this does not appear on the NMI values, since MC has the second best score. This raises a question regarding the appropriateness of this measure to assess community detection performance. IM also finds some small communities, but much less than MC, and the rest of the distribution is more similar to the reference. Compared to the reference and the other algorithms, communities detected by FG and LV have sizes distributed rather uniformly. Interestingly, these two algorithms have very different performances in terms of NMI, so despite the relatively similar sizes of their communities, their community structures are probably very different too. For WT, the size distribution is very close to the reference. Again, this fact alone is not equivalent to a high NMI value, since its performance is substantially lower to IM.

**Table 1.** Algorithms performances, as measured with the Normalized Mutual Information.

| Algorithm | $n = 10^4$ | $n = 10^5, k_{max} = 10^3$ | $n = 10^5, k_{max} = 3 \times 10^3$ |
|---|---|---|---|
| Louvain | 0.80 | 0.78 | 0.80 |
| Fast Greedy | 0.59 | 0.66 | 0.67 |
| MarkovCluster | 0.83 | 0.87 | 0.80 |
| InfoMap | 0.88 | 0.93 | 0.91 |
| Walktrap | 0.77 | 0.79 | 0.78 |

For the embeddedness, MC and WT are clearly different from the reference, displaying a distribution with very few extreme embeddedness values. The small numbers of highly embedded nodes and the fact almost half the nodes have very low embeddedness with MC seems to be linked to the community size distribution. Many of the smallest communities identified by MC are certainly grouped together in the reference, leading to a smaller number of intercommunity links. Compared to the reference, WT does not contain nodes with low embeddedness, whereas it has more nodes with medium embeddedness. In this case, it cannot be related to the community sizes though, since they are comparable to those of the reference. Maybe the lack of low embeddedness nodes can be interpreted as a failure to classify interface nodes, located at the limit of their community and largely connected with other communities. The embeddedness distributions observed for FG and LV are again very similar. They also lack low embeddedness nodes, but not as much as WT. Finally, IM presents the values the most similar to the reference.

When considering the scaled density (Fig. 2), IM, MC and WT are very close to the reference, except IM and MC present very low values for their smallest communities (meaning these are tree-like). For FG and LV, the scaled density is relatively stable, and does not present the slow increase which is characteristic of the reference. This can be interpreted as the fact the communities detected by these algorithms all present the same structure, independently from their size.

The average distances measured on the FG and LV communities are much dispersed and do not follow the evolution observed for the reference. FG, in particular, has a much higher average distance than the reference and the other algorithms. This property is a good indicator of cohesion, so it seems this quality is absent from the communities identified by FG. The remaining algorithms (IM, MC, WT) are very close to the reference. IM displays two outliers though: the average distance is surprisingly high for its smallest and largest communities.

For hub dominance, IM, MC and WT seem to follow the reference, with a positive bias. The fact these algorithms have slightly higher scaled-density and lower hub dominance, relatively to the reference, is consistent with their slightly lower average distance. The inverse observation is valid for the smallest and largest communities detected by IM: sparse and non-centralized communities lead to high average distance. FG and LV once again display similar behaviors, with hub dominance values clearly bellow the reference. When also considering their stable scaled density, this can explain their increasing average distance.

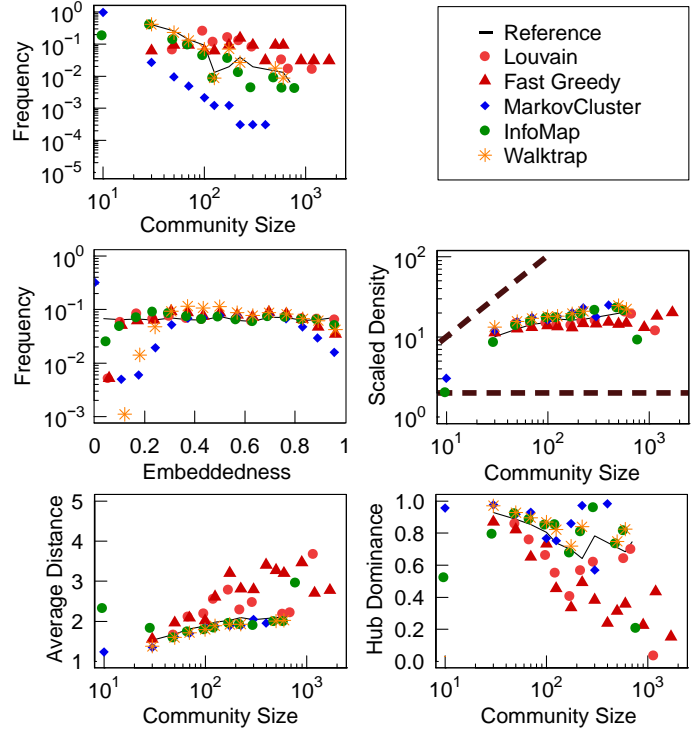

**Fig. 2.** Properties of the detected communities. Each shape/color corresponds to a different algorithm, whereas the reference is represented by a solid line. Points are averages over logarithmic bins of the community size. The dotted lines in the scaled density plot represent the limits of this property, as in Fig. 1.

The topological analysis of the estimated community structures gives a new perspective to the quantitative performance measures. The communities detected by IM, the best algorithm in terms of NMI, are unsurprisingly very close to the reference ones. However, MC, the second algorithm and not far from IM, presents a very different community structure, characterized by much more very small communities. On the contrary, most of the properties of the communities identified by WT are very similar to the reference. It only differs clearly in terms of embeddedness distribution, which is apparently sufficient to rank it only fourth in terms of NMI, relatively far from IM. It thus seems there is no equivalence between a high NMI value and a community structure with properties close to the reference. We conclude both approaches are complementary to perform a relevant analysis of community detection results. It is worth noticing LV and FG, both based on modularity optimization, comparably differ from the reference, which confirms the importance of considering the community definition which characterizes an algorithm.

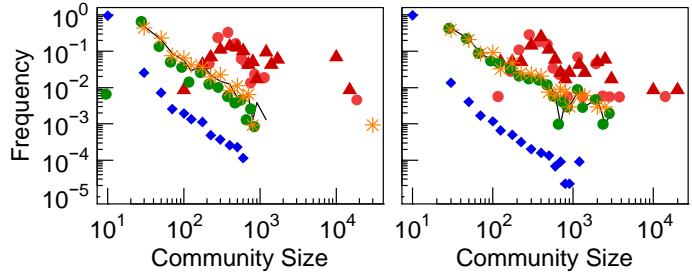

**Fig. 3.** Community size distributions for size $n = 10^5$ networks, with $k_{max} = 1000$ (left) and 3000 (right). Shapes/colors meaning is the same as in Fig. 2, and points are also averages over logarithmic bins of the community size.

On size $n = 100000$ networks, FG and LV find communities much larger than the reference ones, as shown in Fig. 3. For these algorithms, $n_c$ roughly ranges from 100 to 15000, when it goes from 15 to 1000 in the reference. Both are based on modularity optimization, so this might be due to the resolution limit characteristic of this measure [27], which prevents them from finding smaller communities. IM is relatively close to the reference, but not as much as it is on $n = 10000$ networks. WT is also very similar to the reference, but it departs from it by finding a very large community ($n_c \approx 30000$). MC results are relatively similar to those obtained on the $n = 10000$ networks, i.e. it finds many very small communities. In order to separate the effects of network size and density on the algorithms, we generated additional networks with the same size, but maximal degree $k_{max} = 3000$ (instead of 1000). This reduces slightly the overestimation of FG and LV community sizes, whereas MC has roughly the same results. On the contrary, IM and WT properties are excellent, they follow almost perfectly the reference values.

## 5     Conclusion

In this study, we took advantage of recent advances relative to the characterization of community structures in complex networks to tackle two questions: 1) Do artificial networks used as benchmark exhibit real-world community properties? 2) How do community detection algorithms compare in qualitative terms, by opposition to the usual quantitative measurement of their performances. We first applied a variant of the LFR model [6] to generate a set of artificial networks with realistic parameters retrieved from the literature. We studied their properties and concluded some of them are realistic (community sizes, hub dominance), some are only partly realistic (embeddedness, average distance), and others are not realistic at all (scaled density). We then applied on these networks a representative set of five fast community detection algorithms: Fast Greedy, InfoMap, Louvain, MarkovCluster and Walktrap.

It turns out the performance assessed quantitatively through the widely used Normalized Mutual Information (NMI) measure does not necessarily agree with a qualitative analysis of the identified communities. On the one hand, MarkovCluster, ranked second in terms of NMI, actually found an extremely large number of very small communities and almost no large community. On the other hand, the properties of the community structure estimated by Walktrap are very close from the reference ones, but the algorithm comes fourth in terms of NMI, with a score relatively far from MarkovCluster's one. It therefore seems both approaches should be applied to perform a relevant comparison of the algorithms.

Our contributions are as follow. First, we introduced a slight modification to the LFR model, in order to make the embeddedness distribution more realistic in the generated networks. Second, we studied these generated networks in terms of community-centered properties. This complements some previous analyses focusing on network-centered properties such as transitivity or degree correlation [6, 10, 11]. Third, we applied several community detection algorithms on these networks and characterized their results relatively to the same community-centered properties. Previous studies adopted a quantitative approach based on some performance measure [6, 7, 10, 11, 18].

Our work can be extended in various ways. First, it seems necessary to either increase the realism of the LFR model or to define a completely new approach able to generate more realistic networks. Second, by lack of time, we could test only a few algorithms, on a few relatively large networks. A more thorough analysis would consist in using much larger networks, with more repetitions to improve statistical significance. Moreover, applying several algorithms relying on the same definition of the community concept would allow to compare their properties and maybe associate a certain type of community structure to a certain family of algorithms. It could additionally be interesting to use other performance measures than the NMI to assess their relevance with the studied topological properties. Third, it would noticeably be interesting to apply classic network-wise measures to communities (transitivity, degree correlation, centrality, etc.), and to consider additional community specific measures, such as those designed in [13], which seem complementary to the embeddedness, and the concept of community profile [12], although this one looks particularly costly from a computational point of view.

**Acknowledgments.** This project is supported by the Galatasaray University Research Fund.